# Methodological Issues in Building, Training, and Testing Artificial Neural Networks

Stacy L. Özesmi[a]*, Uygar Özesmi[a], and Can O. Tan[b]

[a]Environmental Science Chair, Department of Environmental Engineering, Erciyes University, 38039 Kayseri Turkey

[b]Department of Biology, Middle East Technical University, 06531 Ankara Turkey

* Corresponding author. Environmental Science Chair, Department of Environmental Engineering, Erciyes University, 38039 Kayseri Turkey

Tel: 90 352 437 4937 /32800 Fax: 90 352 437 4404 email: stacy@erciyes.edu.tr

**Abstract**

We review the use of artificial neural networks, particularly the feedforward multilayer perceptron with back-propagation for training (MLP), in ecological modelling. In MLP modeling, there are no assumptions about the underlying form of the data that must be met as in standard statistical techniques. Instead the researchers should make clear the process of modelling, because this is what is most critical to how the model performs and how the results can be interpreted. Overtraining on data or giving vague references to how it was avoided is the major problem. Various methods can be used to determine when to stop training in artificial neural networks: 1) early stopping based on cross-validation, 2) stopping after a analyst defined error is reached or after the error levels off, 3) use of a test data set. We do not recommend the third method as the test data set is then not independent of model development. Many studies used the testing data to optimize the model and training.

Although this method may give the best model for that set of data it does not give generalizability or improve understanding of the study system. The importance of an independent data set cannot be overemphasized as we found dramatic differences in model accuracy assessed with prediction accuracy on the training data set, as estimated with bootstrapping, and from use of an independent data set. The comparison of the artificial neural network with a general linear model (GLM) as a standard procedure is recommended because a GLM may perform as well or better than the MLP. If the MLP model does not predict better than a GLM, then there are no interactions or nonlinear terms that need to be modelled and it will save time to use the GLM. MLP models should not be treated as black box models but instead techniques such as sensitivity analyses, input variable relevances, neural interpretation diagrams, randomization tests, and partial derivatives should be used to make the model more transparent, and further our ecological understanding which is an important goal of the modelling process. Based on our experience we discuss how to build an MLP model and how to optimize the parameters and architecture. The process should be explained explicitly to make the MLP models more readily accepted by the ecological research community at large, as well as to make it possible to replicate the research.

## 1. Introduction

The earliest papers on the use of artificial neural networks in ecology began appearing in the mid 1990's. Reported advantages of artificial neural networks over more traditional methods include: 1) form of relationships need not be specified (no assumptions need to be made about the distribution of data); 2) nonlinear relationships or interactions among variables are easily modelled; 3) performance is usually better than general linear models; 4) complex data patterns can be handled because of their nonlinear nature; 5) ability to generalize to new data. Thus, neural networks have many advantages for ecological studies where data rarely meet parametric statistical assumptions and where nonlinear relationships are prevalent. However, artificial neural networks also have disadvantages: 1) they are computationally intensive; 2) many parameters must be determined with few guidelines and no standard procedure to define the architecture; 3) analyst expertise is required; 4) no global method exists for determining when to stop training and thus overtraining is problematic; 5) sensitive to composition of the training data set; 6) sensitivity of training to initial network parameters; 7) black box models.

Perhaps because it is one of the easiest neural networks to understand, the feedforward multilayer perceptron with back-propagation for training, has been the most commonly used neural network in ecology. More details on how this type of neural network works can be found elsewhere (i.e., Lek and Geugan, 1999 or in texts such as Anderson, 1995; Weiss and Kulikowski, 1991; Bishop, 1995; or Ripley, 1996).

In this article we review the use of the MLP, or feedforward multilayer perceptron with back-propagation for training, in ecological modelling and how it is practiced. Based on our experience we discuss how to build MLP models and how to optimize the parameters and architecture. We make recommendations for best practices, which include the importance of avoiding overtraining, use of an independent test data set, and use of sensitivity analyses, neural interpretation diagrams, input variable relevances, and other methods to open up the black box model. Although in this article we focus on the MLP, some of our recommendations are also relevant to other types of artificial neural networks.

## 2. Literature review

The MLP has been used in ecological studies by Brey et al. (1996) who predicted benthic invertebrate production/biomass ratios, Levine et al. (1996) who classified soil structure from soil sample data, Tan and Smeins (1996) who predicted changes in the dominant species of grassland communities based on climatic input variables, and Poff et al. (1996) who modelled streamflow response based on average daily precipitation and temperature inputs. Paruelo and Tomasel (1997) predicted normalized difference vegetation index (NDVI) used in remote sensing. Phytoplankton production (Scardi, 1996; Scardi and Harding, 1999; Scardi, 2001) and phytoplankton occurrence and succession (Recknagel et al., 1997; Karul et al., 2000) have been modelled with the MLP. Fish abundance based on habitat variables (Baran et al., 1996; Lek et al., 1996), fish yield (Lae et al., 1999), and fish and microhabitat use (Reyjol et al., 2001) have also been modelled. The MLP has been used to predict, based on habitat variables, presence or absence of macroinvertebrates (Hoang et al., 2001), birds (Manel et al. 1999), golden eagle nest sites (Fielding, 1999b), interacting marsh breeding bird nests (Özesmi and Özesmi, 1999), and cyanobacteria (Maier et al., 1998). Bird abundance (Lusk et al., 2001) and macro-invertebrate abundance and species richness (Lek-Ang et al., 1999) has been modelled. The MLP has been used to predict damage to agricultural fields by flamingo (Tourenq et al., 1999) and wild boar (Spitz and Lek, 1999). It has also been used to predict lead concentrations in grasses (Dimopoulos et al., 1999).

*2.1 Criticisms of modelling with MLP*

From a literature review, we saw some problems with the reporting on the use of MLP. Sometimes the modelling process was not clearly described. For example, some research did not report why certain variables were chosen for a final model. Others did not tell how the parameters were set or how the architecture, the number of hidden units, was determined. The number of samples used to train, validate and test the model was not always given.

*2.2 Overtraining*

However, the major problem was overtraining on data or giving vague references on how it was avoided. An exception is Paruelo and Tomasel (1997), who provide a discussion of their experience with overtraining. Unfortunately it seems that often studies do not make sufficient effort to avoid overtraining. One of the biggest disadvantages of using MLP is that there is no perfect method for determining the number of training iterations. There are basically three methods: 1) choose a user defined error level; 2) use an early stopping method such as autotrain (Goodman, 1996); or 3) use the test data. The problem with the first method is that it is difficult to decide on what this error level should be. Often this level is chosen when the error levels off and does not change. The error usually drops until a certain number of iterations where it levels off and does not get much smaller; however at this point the network may be overtrained. With early stopping methods, part of the training data is held out from the training and used to test the model performance. The error goes down on the training data as the training proceeds. The error also initially goes down on the holdout data but then the error level rises again as the model becomes overtrained. While we prefer this method, it requires more data, which often is not available. Another problem with this method is that it doesn't guarantee that the minimum error found is a global minimum rather than a local minimum. If the test data set is used to determine when to stop training, this means that it is not an independent test of the model.

### *2.3 Independent test data*

Another problem we saw with the use of neural networks in ecological research was the lack of independent test data sets. Two main approaches exist for evaluating (testing) model performance (Guisan and Zimmerman, 2000). The first approach is to use a single data set to train and test the model using cross-validation, leave-one-out jack-knife, or bootstrapping. These techniques are used most often when the available data set is small and all the data is needed to train the model. The second approach is to use an independent data set for testing. If the data set is divided into two parts with one part for a test, it is called a split sample approach. However, the independent test is optimal if the two data sets originate from two different sampling strategies.

Some studies used the test data to optimize the model and the amount of training (i.e. Levine et al., 1996 ; Karul et al., 2000; Tourenq et al., 1999; Reyjolet al. 2001). Although this method may give the best model for that set of data it does not give generalizability or improve understanding of the study system.

Studies with independent test data set(s) are rare (i.e. Recknagel et al., 1996; Poff et al., 1996; Paruelo and Tomasel, 1997; Özesmi and Özesmi, 1999). Often the jackknifing (leave-one-out) of the data set is used (i.e. Lae et al., 1999; Manel et al., 1999; Spitz and Lek, 1999) or cross validation (Levine et al., 1996; Lek et al., 1996; Baran et al., 1996; Paruelo and Tomasel, 1997; Manel et al., 1999).

Hirzel et al. (2001) even argue that one independent test is not sufficient but that more tests are needed. However, independent test data is usually in short supply or non-existent in many ecological studies. Therefore they generated a virtual species and used simulated data sets to compare their models. Similar techniques may be possible in other ecological studies as well.

*2.4 Comparison with general linear models*

Many studies have compared the MLP with general linear models (i.e. Baran et al., 1996; Brey et al., 1996; Scardi, 1996; Paruelo and Tomasel, 1997; Fielding, 1999b; Karul et al., 1999; Lae et al., 1999; Manel et al., 1999; Özesmi and Özesmi, 1999). Generally these studies have found that the MLP performs better than general linear models such as multiple linear or logistic regression (i.e. Brey et al., 1996; Baran et al., 1996; Paruelo and Tomasel, 1997; Özesmi and Özesmi, 1999). However the MLP does not always outperform linear techniques (Fielding, 1999b; Manel et al., 1999).

We recommend the comparison of the artificial neural network with a general linear model as a standard procedure because general linear models may perform as well or better than MLP. If the MLP model does not predict better, then there are no interactions or nonlinear terms that need to be modelled and it will save time to use the general linear model. A quick method is to make a neural network without a hidden layer to create the general linear model. If this model performs as well as the MLP with a hidden layer, then there is no need to use the MLP. In addition, by connecting each input to only one hidden unit and then to the output unit a transformation-only neural network model can be made. If the model performance of the MLP with a hidden layer is the same as a transformation-only model, then the variables can simply be transformed to remove nonlinearities in a general linear model (Goodman, 1995).

### *2.5 Opening the black box*

Finally, artificial neural network models should not be treated as black box models but by using various techniques the box can be opened (Scardi, 2001). Available techniques such sensitivity analyses (Lek et al., 1996; Scardi, 1996; Recknagel et al., 1997), input variable relevances and neural interpretation diagrams (Özesmi and Özesmi, 1999), randomization tests of significance (Olden 2000; Olden and Jackson, 2000), and partial derivatives (Dimopoulos et al., 1999; Reyjol et al., 2001) should be used to make the model more transparent. Use of these techniques, which are described below, will further our ecological understanding, which is an important goal of the modelling process.

In sensitivity analyses, the response of the model to each of the input variables is determined by applying a typical range of values to one variable at a time while holding the other variables constant (Lek et al., 1996). The variables that are held constant are set an arbitrary level. The level they are held at influences the results so they can be set at their minimum, first quartile, median or mean, third quartile, and maximum values successively. The resulting plots allow one to examine how the variables influence the model response.

By examining the input variable relevances, we can see how much each input variable contributes to the model (Özesmi and Özesmi, 1999). The relevance of an input variable is the sum square of weights for that input variable divided by the total sum square of weights for all input variables. Variables with high relevances are more important in the model.

Neural interpretation diagrams (NIDs) can be drawn to understand how the model is weighting different input variables and how the input variables interact to give the model response (Özesmi and Özesmi, 1999). NIDs are drawn by scaling the thickness of the lines connecting the units according to the relative values of their weights. Black lines represent positive signals and gray lines represent negative signals. Thus in one diagram, we can look at the thickness of the connections coming out of the input units to see which variables are most important. We can see how the input variables interact and their contribution to model output by looking at the hidden layer. However neural interpretation diagrams are most helpful when the number of units and connections is limited. Diagrams with 20 or more variables are too complicated to gain any insights. For example, twelve variables are typical in analysis of cognitive maps (Buede and Ferrell, 1993).

A randomization test has been developed to assess the statistical significance of connection weights and input variable relevances (Olden, 2000; Olden and Jackson, 2000). In this approach the response variable is randomized, a neural network is constructed using the randomized data, and all the input-hidden-output connection weights (product of the input-hidden and hidden-output weights) are recorded. This procedure is repeated a large number of times to generate a null distribution for each input-hidden-output connection weight. This value is then compared to the actual model value to calculate the significance level. With this randomization test, the neural network can be pruned by eliminating connection weights that have little influence on the network output. With the insignificant connection weights removed, it is easier to interpret how the model makes predictions with a NID. In addition, the randomization test identifies the independent variables that significantly contribute to model prediction.

The partial derivatives of the network output with respect to input variables can be used to show the influence of the variables in the model (Dimopoulos et al., 1999; Reyjol et al., 2001). By plotting the partial derivatives of the network output with respect to an input variable, how the network output changes with increasing values of the input variable can be seen. Somewhat similar to relevances, the importance of the variables in the model can be determined by calculating the sensitivity of the MLP output for the data set with respect to each input variable. The sensitivity is the summation of all the squared partial derivatives for each input variable. By using techniques such as these, MLP models can be easier to interpret and help to improve our understanding of the study system.

**3. ANN modelling process**

Based on our experience we discuss how to build a MLP model and how to optimize the network parameters and architecture. See Tan et al. (2002) for an example of our modeling process where we follow the guidelines and recommendations presented in this paper. Our MLP modelling process generally proceeds as follows. First it is necessary to determine the form of inputs and outputs for the data, the pre and post-processing of the data. Usually the input variables are standardized so that they are all on the same order of magnitude. We have found that standardizing the input variables, to means of zero and units of standard deviations, has consistently led to better results. Then we determine the network parameters such as learn rate, weight range, etc. Next we optimize the architecture, the number of hidden layers and number of hidden units in the hidden layers. Then we optimize the parameters together with the chosen architecture. We use techniques such as neural interpretation diagrams, input variable relevances, and sensitivity analyses to understand how our model is making predictions. Finally we conduct an independent test of the model. When our output is binary and depends on threshold we prefer ROC curves and c-index (which is an estimate of the area under the ROC curve) for assessing model accuracy. See Fielding (1999a) for a discussion of different ways to assess model accuracy.

### *3.1 Network parameters*

Learn rate and weight range are network parameters that influence the performance of the model by affecting the weights. The learn rate and weight range can be set at default values and if the model is unstable made smaller until it stabilizes. We have found that changing the weight range or the learn rate does not result in large changes in model accuracy.

The changes in accuracy are in the same range as those that result from changing the random start, which initializes the weights. Because of the variation in model performance caused by different initial weights, we run all network configurations at least five times using the same predetermined random seeds, produced by a random number generator. Thus we optimize the network parameters and network architecture based on the average of the five random starts.

### *3.2 Architecture optimization*

Next we optimize the architecture, the number of hidden layers and number of hidden units in the hidden layers. Heuristics exist for determining the number of units in the hidden layer(s). However, for each application it is basically a process of trial and error. We systematically run different models to optimize the network architecture. First we create a general linear model (GLM), or a network with no hidden layer. Second we create a transformation only model where each input is connected to only one hidden unit because if the transformation model performs as well as an ANN with a hidden layer then the input variables need only be transformed to remove nonlinearities. Then we create models with one hidden layer having different numbers of hidden units. For a well generalized ANN model, there should be about 10 times as many training data points as there are weights in the network (Bishop, 1995). By using this heuristic, we can set an upper limit on the number of hidden units in the model. We have found with our data that one hidden layer has been as accurate or more accurate than networks with 2 hidden layers (Figure 2). The accuracy of the networks with hidden layers generally increases and then levels off after a certain number of hidden units. When choosing the final architecture, the model with fewer hidden nodes should be chosen, because for two networks with similar errors on the training sets, the simpler one is likely to predict better on new cases (Bishop, 1995).

### *3.3 Avoiding overtraining*

We have had good results with training a MLP and deciding when to stop training by using a holdout set (Özesmi and Özesmi, 1999). However this technique requires more data, which is often in short supply in ecological studies. More recently we have trained MLPs and decided when to stop training by determining when the accuracy leveled off using bootstrapping. We used c-index, which is approximately the area under the Receiver Operating Characteristic (ROC) curve, to assess the accuracy. A c-index of 1 indicates a perfect model and a c-index of 0.5 indicates a model that predicts no better than a random model. We trained the same network to various numbers of epochs and calculated the c-index and corrected c-index with 150 bootstraps. With each bootstrap the model is trained with approximately 66% of the data that is randomly selected from the total set. The model accuracy (c-index) is calculated using that subset of the data and the whole set. The optimistic bias of each bootstrap is determined as the difference between these two accuracies. The corrected c-index is then determined by subtracting the average optimistic bias for all of the bootstraps from the full model c-index. When the corrected c-index levelled off we thought that this would be a well-generalized model. For example, the number of epochs versus c-index is shown in Figure 1. While the c-index continues to increase as the number of epochs increases, the corrected c-index levels off at 70 epochs. This is an indication of overtraining beyond 70 epochs.

*3.4 Independent test data set*

Although it has been stated before many times (i.e. Fielding, 1999a), we cannot overemphasize the importance of an independent testing set. For example, in our work, the c-index of a model based on the training data was 0.746 (Tan et al., 2002). We used 150 bootstraps to estimate how generalizable the model was and the corrected c-index as determined from these bootstraps was 0.663. Finally we tested our model on an independent data set. The c-index was 0.511 or about random. Thus our training data indicated the model was explaining some of the variation in the data. The bootstrapping indicated it was not quite as generalizable but the test data indicated our model was performing about the same as a random model. Thus while very good results might be obtained with training data, and still good results might be indicated by bootstrapping (or some other data splitting technique such as jack-knifing), the real test is the independent data set.

## 4. Conclusions

We recommend that the following information be included in every published research using MLP. These should be included to make the MLP modelling process more transparent and thus more readily accepted by the ecological research community at large, as well as to make it possible to replicate the research. When using standard statistical techniques, the researchers must justify that their data meet the assumptions of those techniques. However in the MLP modeling, there are no assumptions about the underlying form of the data that must be met. Instead the researchers should make clear the process of modelling, because this is what is most critical to how the MLP model performs and how the results can be interpreted. A clear explanation of the modeling processing is necessary including which variables were chosen for the final model and why they were chosen. A description of the form of the input and output variables is needed. For example, input variables are usually standardized so that they are all in the same order of magnitude. An explanation of how the network parameters (learn rate, weight range, momentum) were chosen, and the values that were used in the final model(s) should be stated. How the network architecture was optimized should be included and the number of hidden layers and hidden units that were chosen for the final model(s). The number of samples used to train, validate and test the model should be clearly indicated. This information could be included in an appendix or in a table, but they should be part of every published research.

We have found that one hidden layer is sufficient in our MLP models to achieve high accuracy on the training data and that two hidden layers does not increase this accuracy. The accuracy level levels off after a certain number of hidden units. However to avoid overtraining, the number of training epochs should be limited as well as the number of hidden units in the hidden layer.

Based on the literature review of the use of MLP in ecological research and our own experience, we suggest that more effort be made to interpret the results of the neural network models using techniques such as input variable relevances, sensitivity analyses, neural interpretation diagrams, randomization tests, and partial derivatives. We also recommend that independent test data sets be used for assessing model accuracy, as we found dramatic differences between model performance based on training data, bootstrapping, and use of an independent test data set.

**References**

Anderson, J.A., 1995. An Introduction to Neural Networks. MIT, Cambridge, Massachusetts, p 650.

Baran, P., Lek, S., Delacoste, M., Belaud, A., 1996. Stochastic models that predict trout population density or biomass on a mesohabitat scale. Hydrobiol. 337, 1–9.

Bishop, C.M., 1995. Neural Networks for Pattern Recognition. Oxford University Press, Oxford, p 482.

Brey, T., Jarre-Teichmann, A., Borlich, O., 1996. Artificial neural network versus multiple linear regression: predicting P:B ratios from empirical data. Mar. Ecol. Prog. Ser. 140, 251–256.

Buede, D.M., Ferrell, D.O., 1993. Convergence in problem solving: a prelude to quantitative analysis. IEEE Transactions on Systems, Man, and Cybernetics 23, 746-765.

Dimopoulos, I., Chronopoulos, J., Chronopoulou-Sereli, A., Lek, S., 1999. Neural network models to study relationships between lead concentration in grasses and permanent urban descriptors in Athens city (Greece). Ecol. Model. 120, 157–165.

Fielding, A.H., 1999a. How should accuracy be measured? In: Fielding AH, editor. Machine Learning Methods for Ecological Applications. Kluwer, Dordrecht, p 209-223.

Fielding, A.H., 1999b. An introduction to machine learning methods. In: Fielding AH, editor. Machine Learning Methods for Ecological Applications. Kluwer, Dordrecht, p 1–35.

Goodman, P. H., 1996. NevProp Software, Version 3. University of Nevada, Reno, NV. http://brain.unr.edu/index.php

Guisan, A., Zimmermann, N.E., 2000. Predictive habitat distribution models in ecology. Ecol. Model. 135, 147–186.

Hirzel, A.H., Helfer, V., Metral, F., 2001. Assessing habitat-suitability models with a virtual species. Ecol. Model. 145, 111-121.

Hoang, H., Recknagel, F., Marshall, J., Choy, S., 2001. Predictive modelling of macroinvertebrate assemblages for stream habitat assessments in Queensland (Australia). Ecol. Model. 195, 195–206.

Karul, C., Soyupak, S., Cilesiz, A.F., Akbay, N., Germen, E., 2000. Case studies on the use of neural networks in eutrophication modeling. Ecol. Model. 134, 145–152.

Lae, R., Lek, S., Moreau, J., 1999. Predicting fish yield of African lakes using neural networks. Ecol. Model. 120, 325–335.

Lek, S., Belaud, A., Baran, P., Dimopoulos, I., Delacoste, M., 1996. Role of some environmental variables in trout abundance models using neural networks. Aquat. Living Resour. 9, 23–29.

Lek, S., Guegan, J.F., 1999. Artificial neural networks as a tool in ecological modelling, an introduction. Ecol. Model. 120(2-3), 65–73.

Lek-Ang, S., Deharveng, L., Lek, S., 1999. Predictive models of collembolan diversity and abundance in a riparian habitat. Ecol. Model. 120(2-3), 247-260.

Levine, E.R., Kimes, D.S., Sigillito, V.G., 1996. Classifying soil structure using neural networks. Ecol. Model. 92, 101–108.

Lusk, J.J., Guthery, F.S., DeMaso, S.J., 2001. Northern bobwhite (*Colinus virginianus*) abundance in relation to yearly weather and long-term climate patterns. Ecol. Model. 146, 3–15.

Maier, H.R., Dandy, G.C., Burch, M.D., 1998. Use of artificial neural networks for modelling cyanobacteria Anabaena spp. in the River Murray, South Australia. Ecol. Model. 105(2-3), 257–272.

Manel, S., Dias, J.-M., Ormerod, S.J., 1999. Comparing discriminant analysis, neural networks and logistic regression for predicting species distributions: a case study with a Himalayan river bird. Ecol. Model. 120, 337–347.

Olden, J.D., 2000. An artificial neural network approach for studying phytoplankton succession. Hydrobiol. 436, 131-143.

Olden, J.D., Jackson, D.A., 2000. Illuminating the 'black box': a randomization approach for understanding variable contributions in artificial neural networks. Ecol. Model., .

Özesmi, S.L., Özesmi, U., 1999. An artificial neural network approach to spatial habitat modelling with interspecific interaction. Ecol. Model. 116, 15–31.

Paruelo, J.M., Tomasel, F., 1997. Prediction of functional characteristics of ecosystems: a comparison of artificial neural networks and regression models. Ecol. Model. 98, 173-186.

Poff, N.L., Tokar, S., Johnson, P., 1996. Stream hydrological and ecological responses to climate change assessed with an artificial neural network. Limnol. Oceanogr. 41, 857–863.

Recknagel, F., French, M., Harkonen, P., Yabunaka, K., 1997. Artificial neural network approach for modelling and prediction of algal blooms. Ecol. Model. 96, 11–28.

Reyjol, Y., Lim, P., Belaud, A., Lek, S., 2001. Modelling of microhabitat used by fish in natural and regulated flows in the river Garonne (France). Ecol. Model. 146, 131–142.

Ripley, B.D., 1996. Pattern Recognition and Neural Networks. Cambridge University Press, Cambridge, p 403.

Scardi, M., 1996. Artificial neural networks as empirical models for estimating phytoplankton production. Mar. Ecol. Prog. Ser. 139, 289–299.

Scardi, M., 2001. Advances in neural network modeling of phytoplankton primary production. Ecol. Model. 146, 33–45.

Scardi, M., Harding, L.W.J., 1999. Developing an empirical model of phytoplankton primary production: a neural network case study. Ecol. Model. 120, 213–223.

Spitz, F., Lek, S., 1999. Environmental impact prediction using neural network modelling. An example in wildlife damage. Journal of Applied Ecology 36, 317-326.

Tan, S.S., Smeins, F.E., 1996. Predicting grassland community changes with an artificial neural network model. Ecol. Model. 84, 91–97.

Tan, C.O., Özesmi, S.L., Özesmi, U., Robertson, R.J. Generalizability of artificial neural network models: I. Independent test for predicting breeding success. Ecol. Model. (submitted)

Tourenq, C., Aulagnier, S., Mesleard, F., Durieux, L., Johnson, A., Gonzalez, G., Lek, S., 1999. Use of artificial neural networks for predicting rice crop damage by greater flamingos in the Camargue, France. Ecol. Model. 120, 349–358.

Weiss, S.M., Kulikowski, C.A., 1991. Computer Systems that Learn. Morgan Kauffman, San Mateo, CA, p 223.

## Figure Captions

Figure 1. Average and standard deviation of c-index and corrected c-index versus the number of epochs the MLP model was trained.

Figure 2. MLP model accuracy on the training data versus the number of hidden units in one hidden layer (top curve) and two hidden layers (bottom curve).

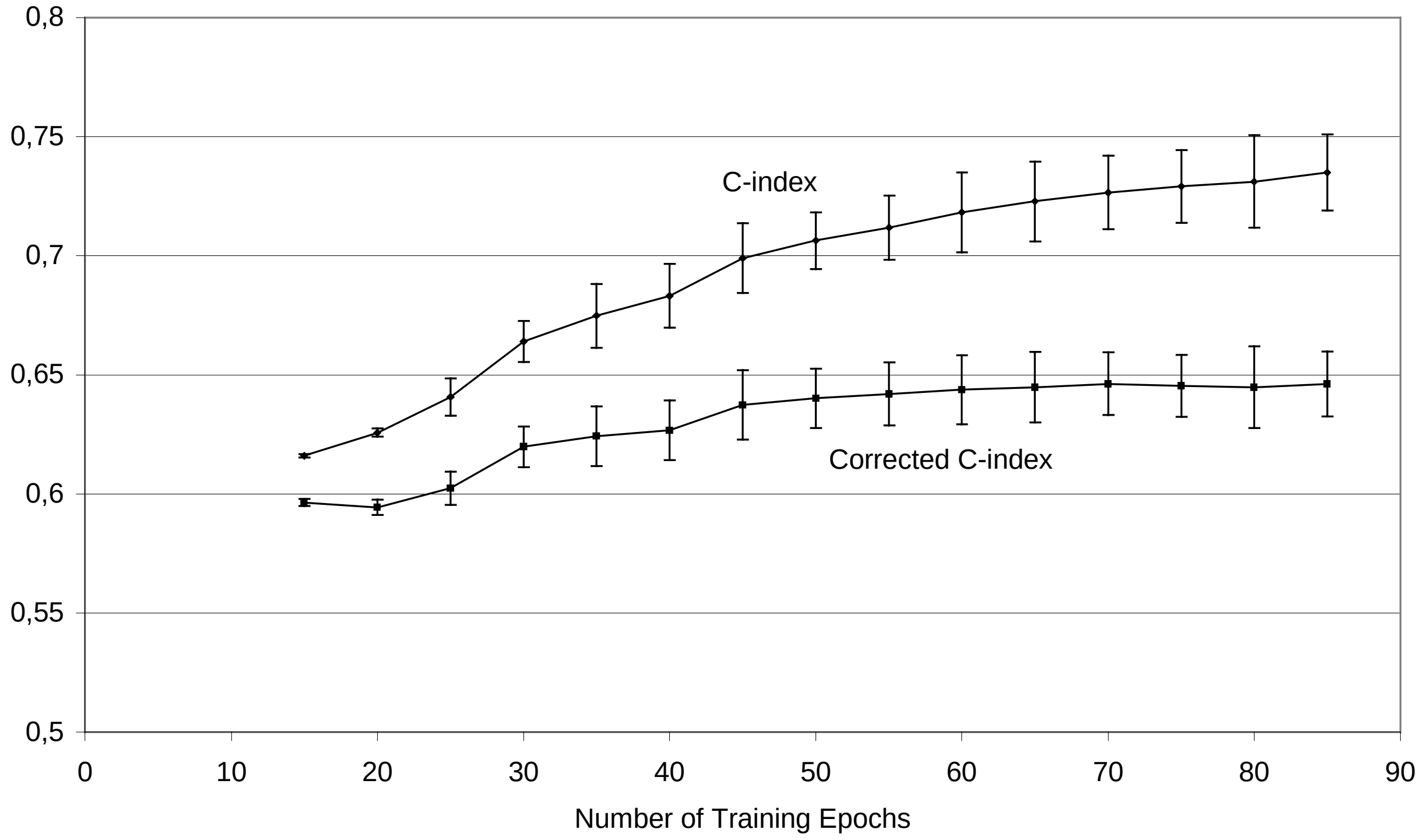

Figure 1.

Figure 2.

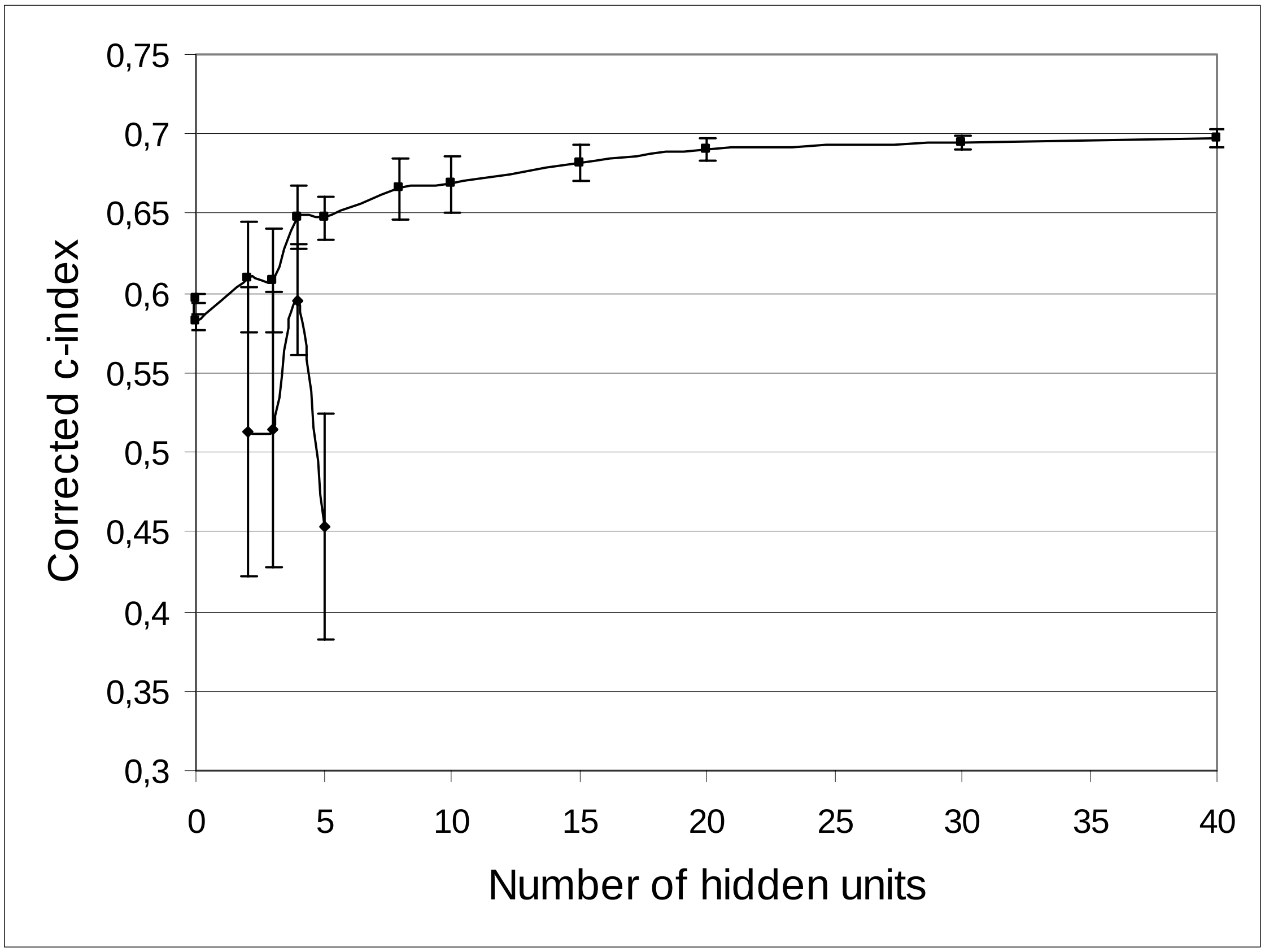